\documentclass[12pt]{article}
\usepackage[latin1]{inputenc}
\usepackage{cite}
\usepackage{amsmath}
\usepackage{amsfonts}
\usepackage{amssymb}
\usepackage{geometry}
\usepackage{amssymb,epsfig}
\usepackage{hyperref}

\makeatletter
\renewcommand\section{\@startsection {section}{1}{\z@}%
                                 {-3.5ex \@plus -1ex \@minus -.2ex}
                                   {2.3ex \@plus.2ex}%
                                   {\normalfont\large\bfseries}}
\renewcommand\subsection{\@startsection{subsection}{2}{\z@}%
                                   {-3.25ex\@plus -1ex \@minus -.2ex}%
                                     {1.5ex \@plus .2ex}%
                                     {\normalfont\bfseries}}
\renewcommand\subsubsection{\@startsection{subsubsection}{3}{\z@}%
                                   {-3.25ex\@plus -1ex \@minus -.2ex}%
                                     {1.5ex \@plus .2ex}%
                                     {\normalfont\itshape}}
\makeatother

\def\pplogo{\vbox{\kern-\headheight\kern -29pt
\halign{##&##\hfil\cr&{\ppnumber}\cr\rule{0pt}{2.5ex}&\ppdate\cr}}}
\makeatletter
\def\ps@firstpage{\ps@empty \def\@oddhead{\hss\pplogo}%
  \let\@evenhead\@oddhead 
}
\def\maketitle{\par
 \begingroup
 \def\thefootnote{\fnsymbol{footnote}}
 \def\@makefnmark{\hbox{$^{\@thefnmark}$\hss}}
 \if@twocolumn
 \twocolumn[\@maketitle]
 \else \newpage
 \global\@topnum\z@ \@maketitle \fi\thispagestyle{firstpage}\@thanks
 \endgroup
 \setcounter{footnote}{0}
 \let\maketitle\relax
 \let\@maketitle\relax
 \gdef\@thanks{}\gdef\@author{}\gdef\@title{}\let\thanks\relax}
\makeatother

\def\prslash{{\partial\mkern-9mu/}}    

\numberwithin{equation}{section}

\newcommand{\be}{\begin{equation}}
\newcommand{\bea}{\begin{eqnarray}}
\newcommand{\ee}{\end{equation}}
\newcommand{\eea}{\end{eqnarray}}

\newcommand{\LHL}{\Lambda_{\textrm{HL}}}
\newcommand{\LUV}{\Lambda_{\textrm{UV}}}

\renewcommand{\a}{\alpha}
\renewcommand{\b}{\beta}
\newcommand{\g}{\gamma}
\renewcommand{\r}{\rho}
\newcommand{\s}{\sigma}

\begin{document}
 
\setcounter{page}0
\def\ppnumber{\vbox{\baselineskip14pt
}}
\def\ppdate{\footnotesize{}} \date{}

\author{Maxim Pospelov$^{1,2}$, Carlos Tamarit$^{2}$\\
[7mm]
{\normalsize $^1$\it Department of Physics and Astronomy, University of Victoria,}\\
{\normalsize \it Victoria, BC, V8P 5C2, Canada}\\
{\normalsize  $^2$\it Perimeter Institute for Theoretical Physics, Waterloo, ON, N2L 2Y5, Canada}\\
[3mm]
{\tt \footnotesize  pospelov at uvic.ca, ctamarit at perimeterinstitute.ca }
}

\title{\bf Lifshitz-sector mediated SUSY breaking 
\vskip 0.5cm}
\maketitle

\begin{abstract} \normalsize
We propose a novel mechanism of SUSY breaking by coupling a 
Lorentz-invariant supersymmetric matter sector to non-supersymmetric gravitational interactions 
with Lifshitz scaling. 
The improved UV properties of Lifshitz propagators moderate the otherwise uncontrollable ultraviolet divergences
induced by gravitational loops. This ensures that both the amount of induced  Lorentz violation and 
SUSY breaking in the matter sector are controlled by $\LHL^2/M_P^2$, the ratio of the Ho\v rava-Lifshitz cross-over scale
$\LHL$ to the Planck scale $M_P$. This ratio can be kept very small,  providing a novel way of 
explicitly breaking supersymmetry without reintroducing fine-tuning. We illustrate our idea by considering
a model of scalar gravity with Ho\v rava-Lifshitz scaling coupled to a supersymmetric Wess-Zumino matter sector, in which we compute the two-loop SUSY breaking  corrections to the masses of the light scalars due to the 
 gravitational interactions and the heavy fields. 
\end{abstract}
\bigskip
\newpage

\section{Introduction\label{sec:intro}}

Supersymmetry (SUSY) is a vastly studied framework, motivated by its ability to solve the gauge hierarchy problem. 
The latter belongs to the class of ``technical naturalness'' problems, 
and  is usually formulated in terms of the quadratic divergences
plaguing the Higgs mass term in the effective potential. In the absence of 
 protection mechanisms, based for example on symmetries, the physical mass of the Higgs field is 
naturally driven towards the cutoff of the theory unless some extreme fine-tuning
of parameters is invoked. Since the quadratic divergences are scheme dependent --absent, for example, in dimensional regularization-- and hence unphysical, one may want to re-state the same problem 
in an alternative way: the Higgs mass is sensitive to generic New Physics in the form of heavy states,
which may couple to the Higgs field either directly or via other gauge and matter fields of the 
Standard Model (SM). For example, generic heavy states of mass $M$ 
would normally  give rise to finite threshold contributions to the 
Higgs mass of the form $\frac{M^2}{16\pi^2}$, which again tend to drive the physical 
mass towards unacceptably large values so that an {\em ad hoc} fine adjustment of the Higgs mass is required. 

Supersymmetry solves this problem by automatically forcing the cancellation of threshold contributions between fermionic and bosonic degrees of freedom in the ultraviolet (UV). However, since SUSY is not  realized  exactly in Nature, there must be new fields and interactions responsible for its breaking. If the main phenomenological motivation for SUSY is 
to be kept, the SUSY breaking mechanisms should not reintroduce the dangerous 
quadratic divergences (or threshold contributions). The most common approach to this 
problem is to assume that SUSY is spontaneously broken at some energy scale, 
so that nonlinearly realized SUSY still forbids quadratic divergences, and the finite threshold corrections to the Higgs mass are proportional to $m_b^2-m_f^2$, 
the difference between the squared masses of bosons and fermions after the SUSY breaking. 
These considerations, together with the naturalness requirement 
of no tuned  cancellations between  threshold corrections and the bare Higgs mass itself, set the expectations 
for finding supersymmetric partners  in the TeV range. (This logic applies  
at least to those superpartners that have significant coupling to the Higgs field.)

If supersymmetry is broken by hard interactions, one expects the comeback of dangerous quadratic divergences and threshold corrections. For example, if the top and stop Yukawa couplings were different even by a tiny amount
$\Delta y_t$ in the whole dynamical range of energies, 
a quadratic divergence would be resurrected, signaling sensitivity to the highest energy scale:  
$\delta m_H^2 \propto y_t\Delta y_t \times \LUV^2$. 
One possibility is that SUSY could be broken by higher dimensional operators involving some inverse power of a large mass scale $\LUV$,  suppressing quantum corrections. A chief example in this category is given by  non-supersymmetric gravitational interactions, with $\LUV$ identified 
with the Planck mass $M_P$ (or, equivalently, supergravity with $m_{\rm gravitino}\to M_P$). 
However, not only there will be nonzero quadratic and even higher power divergences, but 
the finite threshold corrections due to possible heavy states of mass $M$ will scale as $\frac{M^4}{\LUV^2}$, hence becoming unacceptably large for $M\sim\LUV$. Thus, either the new interactions would need to become supersymmetric, perhaps at or below some intermediate scale $\sim (\LUV m_H)^{1/2}$, or a new mechanism for naturalness would need to be invoked at a scale below $\LUV$. 

An interesting exception to these otherwise quite generic arguments is a possible change in the 
dynamics of the New Physics (NP) that renders the power-counting based arguments above not valid. 
This can happen if the interactions in the NP sector, aside from being proportional to inverse powers of $\LUV$,
stop growing in the UV at some additional intermediate scale $\Lambda_{\rm inter} \ll\LUV,M_P$. 
Then it is possible for threshold corrections to the Higgs mass to
pick up suppression factors of the form $\Lambda_{\rm inter}/\LUV$. 
This can happen in theories where $\Lambda_{\rm inter}$ serves as a cross-over scale 
for the dispersion law of elementary excitations, changing from $E=|\vec{p}|$ below this scale to 
a higher power of $|\vec{p}|$ above it. This is precisely the situation in
Ho\v rava-Lifshitz type (HL) theories, where the propagators of particles from the Lifshitz sector have a characteristic form
\begin{align}
\label{eq:Horavaprop}
 \frac{i}{E^2-|\vec{p}|^2-\LHL^{2-2z}|\vec{p}|^{2z}-\delta m^2},
\end{align}
with some $z>1$ and the cross-over momentum scale $\LHL$. 
As pointed out by Ho\v rava, such 
theories with $z\geq 3$ can be a promising candidate for a renormalizable theory of gravitational interactions 
 \cite{Horava:2009uw}. As is obvious from the form of the propagator (\ref{eq:Horavaprop}), Lorentz symmetry is 
broken around the scale $\LHL$, which may present additional phenomenological challenges to such models.
However, if their Lorentz-violating phenomenology can be brought under control, then one might benefit from much 
faster UV convergence in loop diagrams involving the propagator in eq.~(\ref{eq:Horavaprop}).

Indeed, we suggest  to consider the case in which gravitational interactions produce hard SUSY breaking, with $\LUV$ identified with 
$M_P$.  Given the improved behavior of the gravity propagators, divergences will not only be milder but, together with the threshold corrections associated with heavy states, will involve nonzero powers of $\frac{\LHL}{M_P}$. Demanding the absence of large SUSY-breaking corrections to the
 masses of the SM superpartners suggests then a hierarchy of scales, namely ${\LHL}\ll{M_P}$. Intriguingly, the same hierarchy is 
required to suppress the amount of Lorentz violation (LV) transmitted to the matter sector via gravitational loops \cite{Pospelov:2010mp}.

In this paper we set to evaluate the plausibility of this picture by 
computing threshold corrections in what appears to be the simplest model capturing the essentials of the dynamics discussed above.
Specifically, we consider  a supersymmetric Wess-Zumino sector with light and heavy fields coupled to scalar gravity with Lifshitz scaling,
and calculate loop corrections to the boson mass of the light superfield. It will be shown that  the SUSY-breaking threshold corrections to the mass of the light scalar involving the heavy mass scale $M$, which appear at two loops,  are indeed suppressed by powers of $\frac{\LHL}{M_P}$ in the limit of small $\LHL$  and can be made phenomenologically acceptable even for $M\sim M_P$. Although quadratic divergences reappear, they are also suppressed by powers of $\frac{\LHL}{M_P}$. By starting from exact supersymmetry 
in the limit $M_P\to \infty$, we can also increase the degree of protection against LV in the matter sector \cite{GrootNibbelink:2004za,Bolokhov:2005cj}.

The paper is organized as follows: in the next section we will 
elaborate on our proposal in more detail, and discuss known consequences of 
introducing LV into SUSY theories. In Section 3 we introduce the simplest test-ground
for our proposal: a toy supersymmetric model 
coupled to scalar gravity with Lifshitz propagators. We also derive the necessary Feynman rules. Section 4 is the main part 
of our paper, where the two-loop corrections to the light scalar masses are calculated in the $\LHL/M_P$ expansion. 
We reach our conclusions in section 5. Appendix  \ref{app:integrals} contains technical details on the 
evaluation of two-loop integrals with some Lifshitz propagators 
using  dimensional regularization. 


\section{Taming LV by scale separation\label{sec:taming_LV}}

Whichever additional theoretical flexibility Lorentz violation may offer, it must confront 
extremely precise experimental tests of this symmetry. Indeed, neither
studies of high-energy cosmic rays  and associated phenomena 
nor the most precise low-energy measurements of atomic and 
particle systems have produced any credible hints of the departure 
from Lorentz invariance (for some reviews on the subject see {\em e.g.} Refs. \cite{Pospelov:2004fj,Jacobson:2005bg}).
 Given that the straightforward  classification of 
Lorentz-violating operators \cite{Colladay:1996iz} shows that they are at least of mass dimension 3, or mass dimension 4 in the case of
CPT-preserving backgrounds, one may wonder if  a
high-energy theory can be made Lorentz-violating in a phenomenologically consistent way. Actually, if Lorentz invariance is 
completely broken at some high-energy scale ({\em e.g. $M_{P}$}), the rules of effective 
theories and the radiative transfer of LV from one field to another would virtually 
guarantee large amounts of LV at low energy. In particular, one would expect 
dimension 3 and 4 Lorentz-violating operators proportional to the first and zeroth power of that high scale. 
These are huge amounts of LV, that are clearly inconsistent with 
modern limits, which require that the differences in the speed of propagation for different species must not 
exceed $\Delta c/ c \sim 10^{-22}$. 

Therefore it is clear that {\em if}  LV is to be a property of  high energy physics, a mechanism should be found ensuring that 
the corresponding LV at low energy  is sufficiently suppressed by powers of some IR scale over the appropriate UV scale, such as $M_{P}$. 
(A classification of all possible operators of this type for the Standard Model can be found in \cite{Bolokhov:2007yc}). 
Scenarios involving  strong interactions that, together with an
appropriate sign for the anomalous dimensions of Lorentz-violating operators, would suppress their contributions
at low energy, have been advocated on several occasions (see {\em e.g.} Refs. \cite{Anber:2011xf,Bednik:2013nxa}). 
We do not pursue this solution here because we would like to stay on fully perturbative grounds. Instead, 
we consider mechanisms of suppressing
LV for the SM fields relying on scale separation as well as on supersymmetry and the protection it offers against large 
radiative transfer, building on the ideas of Refs. \cite{Pospelov:2010mp} and \cite{GrootNibbelink:2004za,Bolokhov:2005cj}.

The main idea of \cite{Pospelov:2010mp} is that, if LV is sourced by high-energy Lifshitz behavior, then in order to tame LV in the matter sector at low energy one should 
{\em i.} limit the Lifshitz behavior exclusively to the gravity sector, and {\em ii.} ensure that the gravitational 
and HL scale are widely separated, ${\LHL}\ll{M_P}$. As a consequence of the first point, the different 
species of the SM with  different spins ({\em e.g.} gauge and Higgs bosons) 
acquire deviating, loop-induced Lorentz-violating corrections to the limiting propagation speed, so that
\be
\frac{\Delta c}{c} \sim O(1) \times \frac{\LHL^2}{16 \pi^2  M_P^2} \log\left( \frac{\LUV^2}{\LHL^2} \right).
\ee
This result implies that Lorentz-violating corrections can be brought under phenomenological control 
for $\LHL$ below some intermediate scale of $\sim 10^{10}$ GeV. Concrete implementation of this scale 
separation proposal within Ho\v rava gravity meets some difficulties due to the non-Lifshitz spin=1 sector of the 
gravitational interactions, which needs to be supplied with additional terms beyond those in the original gravitational action
with anisotropic scaling  \cite{Horava:2009uw}. It may look somewhat artificial that matter and 
gravity should have different scalings of the 
propagators in the UV, but the alternative, Lifshitz-type matter, requires enormous fine-tuning 
because of simple SM loops being able to induce $O(1\%)$ nonuniversality in the propagation speed 
for different species (see {\em e.g.} \cite{Iengo:2009ix}). Further insights on calculations of loop corrections in 
HL theories can be found in Ref. \cite{Kimpton:2013zb}, while the current status of developments in these 
theories can be found in these works: \cite{Blas:2010hb,Mukohyama:2010xz,Griffin:2011xs,Sotiriou:2010wn}. 

As is clear from the above discussion, the main difficulty in implementing the  proposal \cite{Pospelov:2010mp} is the lack of any 
argument justifying why LV should not be present in the matter sector at all to begin with. 
A possible resolution of this problem can be found within the framework of the 
supersymmetric Standard Model (MSSM), where it was shown \cite{GrootNibbelink:2004za,Bolokhov:2005cj} 
that in the limit of exact SUSY there is an automatic localization of LV to higher dimensional operators 
(dim=5, 6 etc). Once  SUSY becomes broken, one finds that lower dimensional operators are induced,
\be
 {\cal O}_{LV,MSSM}^{\rm dim =6}  \to  {\cal O}_{LV,SM}^{\rm dim =4}  \times (m_{b}^2-m_f^2).
\ee
As a result, again, the SM can be protected from LV by the wide scale separation between the SUSY breaking mass parameters and 
scales normalizing dimension 6 operators, or $m_{\rm~ SUSY~breaking} \ll M_P$. However, lifting these ideas 
to the level of supergravity was never attempted, and it is not known whether this is possible. 

In this paper we propose to combine together both ideas of scale separation in HL gravity  and protection against 
LV by SUSY. Instead of trying to supersymmetrize  HL gravity, we propose to leave this sector 
completely nonsupersymmetric, and make the matter sector obey exact SUSY in the $M_P\to \infty$ limit. 
The self-consistency of this scenario has to be checked by investigating the transfer of the hard breaking of 
SUSY in the gravitational HL sector to the matter sector. If the results for the MSSM SUSY breaking parameters 
were to come out unsuppressed by $\LHL$, {\em e.g.} $m_b \propto \LUV^2/M_P$, then there would be no benefits and no 
real grounds for adopting SUSY in the matter sector to begin with, since naturalness would be lost. If on the other hand we were to find that the amount 
of SUSY breaking is to be controlled by $\LHL$, one could bring both SUSY and LV breakings under control,
have a candidate theory for a renormalizable theory of gravity, and address the hierarchy problem in the SM sector. It is the latter option
that seems to hold, as shown in the rest of the paper by performing explicit calculations in a toy model capturing the essential features of the ideas discussed above.

\section{WZ model coupled to scalar gravity with Lifshitz scaling\label{sec:WZmodel}}

In order to  study the amount of SUSY breaking in the matter sector induced by the 
HL gravitational interactions, we build the simplest model that has all the required ingredients. 
Specifically we choose the following matter content: 
\begin{itemize} 

\item   A chiral matter superfield  $\phi$  (which will also denote its scalar component)
that should remain light in the IR, a prototype for a generic MSSM superfield. 

\item  A very heavy matter superfield 
$\Phi$ that represents generic new physics at a scale $m_\Phi = M$, that we can take as high as 
the Planck scale. This superfield has a Yukawa-type interaction with $\phi$, which serves as a prototype for the coupling of 
MSSM fields to new physics at UV scales. 

\item One light real scalar field $\chi$ ({\em not} a superfield!) with Lifshitz scaling. We choose it to couple to the 
trace of the stress-energy tensor for the matter fields with a $\sim 1/M_P$ coefficient. Therefore $\chi$, a scalar graviton,  is the prototype for a more realistic version of HL gravity. 

\end{itemize} 

Our main goal is to study the sensitivity of the mass of the bosonic component of $\phi$ on the heavy threshold 
$M$, when interactions with the explicitly nonsupersymmetric gravitational Lifshitz sector are turned on. 
Phrasing the hierarchy problem in the language of threshold effects
saves us from the regularization ambiguities 
normally associated with the hard cutoff schemes.

We remind the reader of the Lagrangian for a Wess-Zumino model with superpotential $W$
in flat space.  For a collection of chiral multiplets  labeled here by an index $i$, each including a scalar field $\varphi_i$ 
and a Majorana fermion $\psi_i$, one has 
\begin{align*}
  {\cal L}_{m}=\partial_\mu\varphi^*_i\partial^\mu\varphi_i+\frac{i}{2}\bar{\psi_i}\prslash \psi_i-\left(\frac{\partial W}{\partial \varphi_i}\right)^\dagger\frac{\partial W}{\partial \varphi_i}
  -\left\{\frac{1}{2}\frac{\partial^2 W}{\partial\varphi_i\partial\varphi_j}\bar\psi_i\frac{(1-\gamma_5)}{2}\psi_j+\text{h.c.}\right\},
\end{align*}
where repeated indices are summed. In practice, as said above we consider two chiral multiplets 
with scalars $\phi,\Phi$ and fermions $\psi,\Psi$. Their masses and Yukawa interaction come from the simplest 
renormalizable superpotential:
\begin{align*}
 W(\phi,\Phi)=\frac{1}{2}m\phi^2+\frac{1}{2}M\Phi^2+\lambda\phi \Phi^2.
\end{align*}
The coupling of matter fields to scalar gravity mediated by a real scalar field $\chi$ is given by the interaction
\begin{align*}
 {\cal L}_{m\chi}=-\frac{\kappa}{2\sqrt{2}}\,\chi\, {{T_{m}}^\mu}_\mu=-\frac{\kappa}{2\sqrt{2}}\chi\left(2\partial_\mu\varphi^*_i\partial^\mu\varphi_i+\frac{i}{2}\bar{\psi_i}\prslash \psi_i-4 {\cal L}_m\right),
\end{align*}
where ${T_m}^{\mu\nu}$ is the energy-momentum tensor of the matter sector, and $\kappa$ is up to a coefficient the inverse 
of the  Planck mass, $\kappa=\sqrt{32\pi G_N}=\sqrt{32\pi c\hbar}M_{P}^{-1}$. 
This interaction is the same that would be obtained by coupling the WZ model to ordinary linearized gravity and
identifying $\chi$ with the trace of the metric fluctuation
\begin{align*}
 h_{\mu\nu}=\frac{1}{\sqrt{2}}\eta_{\mu\nu}\chi.
\end{align*}

For the kinetic term of the scalar graviton, we consider one giving rise to a propagator of Lifshitz type with a scale $\Lambda$
(we remove the subscript "HL" for concision in the following):
\begin{align*}
 {\cal L}_{kin, \chi}=\frac{1}{2}\chi\,\left(\partial^2-\frac{(\vec{\partial}\cdot\vec{\partial})^3}{\Lambda^4}-\delta m^2\right)\chi.
\end{align*}
Note that we have added a mass $\delta m$, since the mass of the scalar graviton is not protected by gauge symmetry. 
In keeping with regarding this model as a toy model for massless gravitational interactions, we are interested in the limit of $x\equiv\frac{\delta m^2}{M^2}\rightarrow0$, in which case
$\delta m^2$ can be seen as an IR regulator. We expect the final result for the threshold corrections to be free of IR divergences, which will serve as a consistency check for the calculation.

\begin{figure}
 \includegraphics[width=1\textwidth]{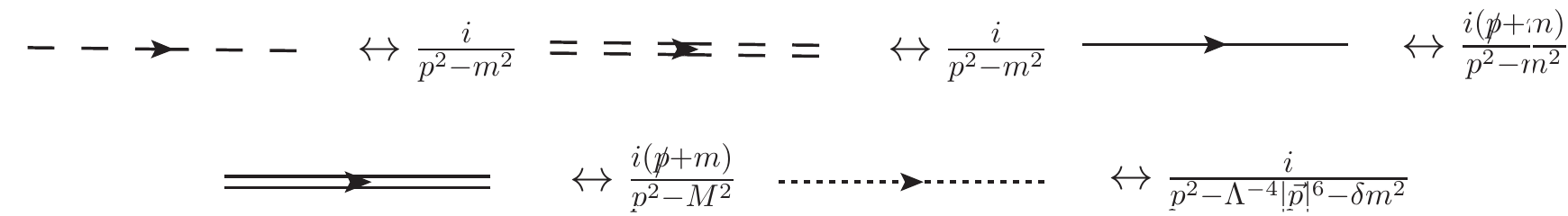}\vskip15pt
 \includegraphics[width=1\textwidth]{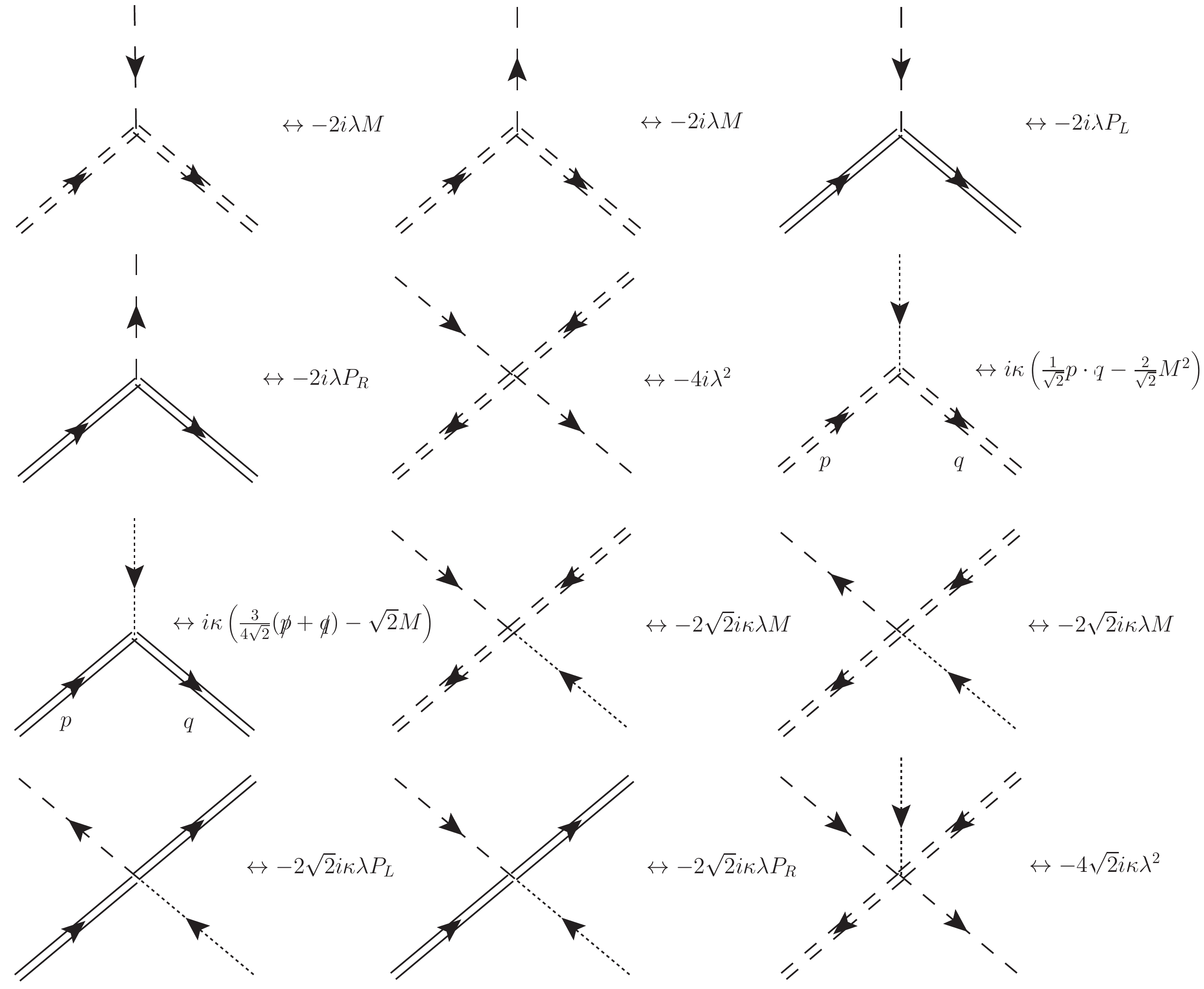}
\caption{
\label{fig:Feynman_rules}Feynman rules relevant for the calculation of the diagrams in fig.~\ref{fig:diagrams}. The fields from the heavy chiral multiplet are denoted with double lines. Fermion lines
are solid, and scalar graviton's are dotted}
\end{figure}

The Feynman rules relevant for our computation are shown in fig.~\ref{fig:Feynman_rules}. 
The fields from the heavy chiral multiplet are denoted with double lines. Fermion lines
are solid, and scalar graviton's are dotted. 

\section{Threshold corrections involving the heavy masses}

As already explained, we are interested in evaluating the light scalar's SUSY-breaking threshold corrections induced by the Lifshitz
dynamics at high energy scales and involving the masses of the heavy fields.  In this way we will be able to test whether these scalar-gravity-induced 
contributions are under control for a suitable choice of the Lifshitz scale $\Lambda$.

\begin{figure}
 \includegraphics[width=1\textwidth]{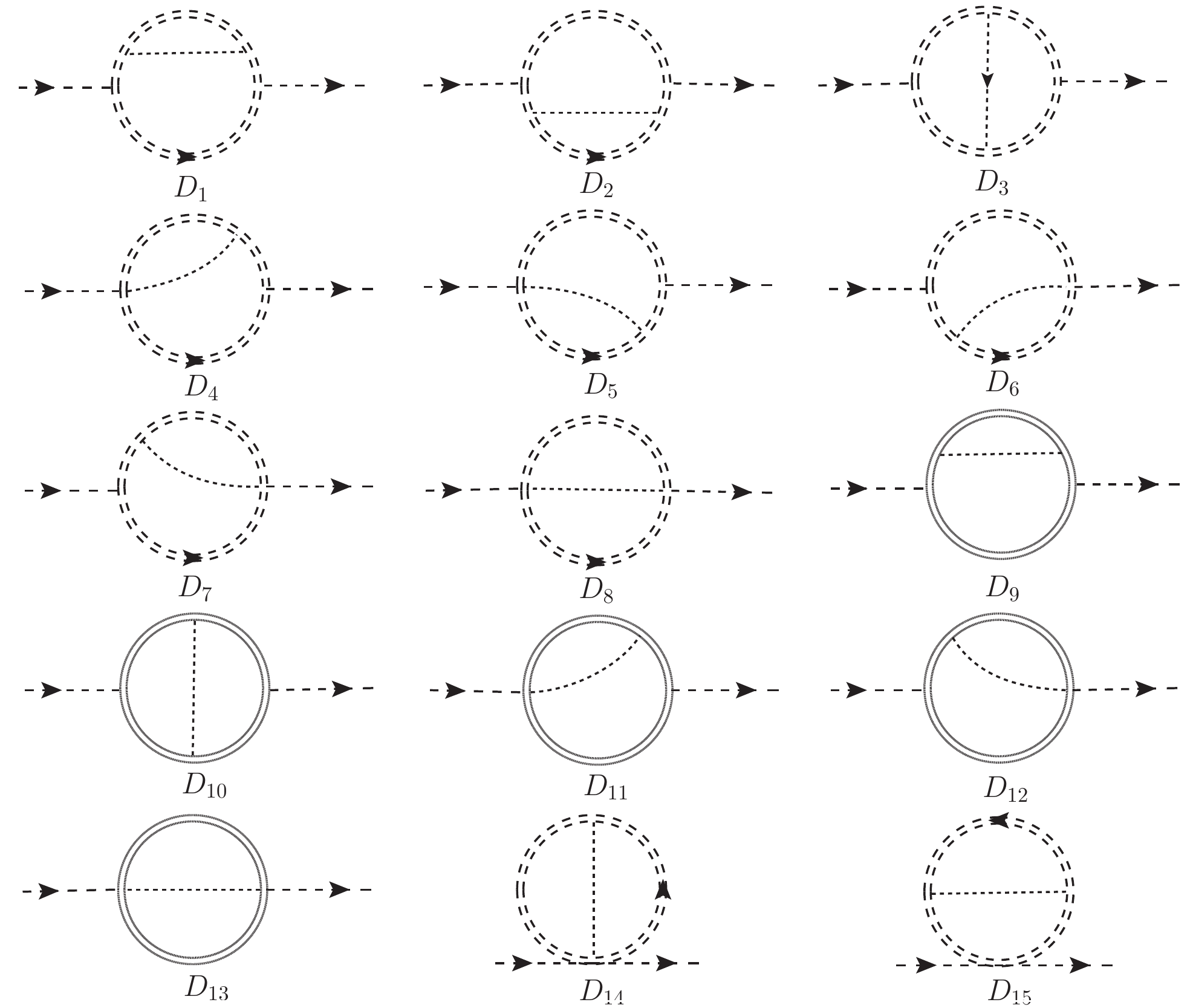}
 \caption{\label{fig:diagrams}Two loop diagrams dominating the threshold corrections to the light scalar's mass due to the heavy fields and the scalar graviton.}
\end{figure}

Since SUSY guarantees the nonrenormalization of the potential in the absence of the scalar graviton interactions
($\kappa\to 0$ limit), we need to calculate diagrams involving the latter. 
At one loop, all possible diagrams have no lines corresponding to heavy fields and thus 
will not give rise to any dependence on the heavy mass $M$ and will be ignored. The dominant diagrams contributing at two-loops are shown in fig.~\ref{fig:diagrams}. Note that we have not included diagrams with scalar gravitons attached to the external light scalar legs through 3-point vertices. 
This is  because such  diagrams become proportional to the IR parameters, either the external momenta or 
the light mass $m$, and are thus subdominant.  For similar reasons it is safe to ignore 
the counterterm diagrams corresponding to the one-loop divergences. 
For convenience, the external momentum can be put to zero for all diagrams. 

We have computed the diagrams using dimensional regularization in $D=4-2\epsilon$ dimensions, using the Feynman rules in fig.~\ref{fig:Feynman_rules}. All integrals can be reduced to the form 
\begin{align}
 I[\alpha,\beta,\gamma,\rho,\sigma]=\int \frac{d^Dk d^Dl}{(2\pi)^{2D}}\frac{(l_0)^\rho (|\vec{l}|)^\sigma}{[k^2-M^2]^\alpha[(k+l)^2-M^2)]^\beta[(l^2-\Lambda^{-4}|\vec{l}|^6-\delta m^2)]^\gamma},
 \label{eq:masterint}
\end{align}
with no $k$ dependence in the numerator. This is because the integrands obey the following relations,
\begin{align*}
 \nonumber&k^2 P[\a,\b,\g]=P[\a-1,\b,\g]+M^2 P[\a,\b,\g],\\
 &(l\cdot k)P[\a,\b,\g]=\frac{1}{2} P[\a,\b-1,\g]-\frac{l^2}{2}P[\a,\b,\g]-\frac{1}{2}P[\a-1,\b,\g],
\end{align*}
which can be applied recursively. $P[\alpha,\beta,\gamma]$ here is defined as
\begin{align*}
 P[\alpha,\beta,\gamma]\equiv\frac{1}{[k^2-M^2]^\a}\frac{1}{[(k+l)^2-M^2]^\b}\frac{1}{[l^2-\Lambda^{-4}|\vec{l}|^6-M^2]^\g}.
\end{align*}

In doing so, one arrives at some integrals with $\alpha=-1$, which can be obtained from $I[0,\beta,\gamma,\rho,\sigma]$
using the identity
\begin{align*}
 I[-1,\b,\g,\r,\s]=I[0,\b,\g,\r+2,\s]-I[0,\b,\g,\r,\s+2]-M^2 I[0,\b,\g,\r,\s].
\end{align*}
A further symmetry property simplifying the calculations is
\begin{align*}
I[\a,\b,\g,\r,\s]=I[\b,\a,\g,\r,\s].
\end{align*}
All the integrals needed for the calculation are obtained in appendix \ref{app:integrals}, where analytic formulae are given for the dominant contributions in the limit of small $\Lambda$ and $x=\frac{\delta m^2}{M^2}$. 
These limits suit our goal of checking whether a small value of $\Lambda$ is able to suppress the contributions to soft masses due to loops of very heavy fields with masses $M\gg\Lambda$.

After using the above properties, the diagrams of fig.~\ref{fig:diagrams} have the following expressions in terms of the family of integrals $I[\a,\b,\g,\r,\s]$:
\begin{align*}
 D_1&=D_2=\frac{1}{2} i \kappa ^2 \lambda ^2 M^2 \left(4 M^4 I[1,3,1,0,0]-5 M^2 I[0,3,1,0,0]\right.\\
 &-4 M^2 I[1,2,1,0,0]-4 M^2 I[1,3,1,0,2]+4 M^2 I[1,3,1,2,0]+2 I[0,2,1,0,0]\\
 &+I[0,3,1,0,2]-I[0,3,1,2,0]+I[1,1,1,0,0]+2 I[1,2,1,0,2]-2 I[1,2,1,2,0]\\
 &\left.+I[1,3,1,0,4]-2 I[1,3,1,2,2]+I[1,3,1,4,0]\right),\\
 D_3&=\frac{1}{2} i M^2 \kappa ^2 \lambda ^2 \left(I[0,2,1,0,0]+2 I[1,1,1,0,0]-4 M^2 I[1,2,1,0,0]+2 I[1,2,1,0,2]\right.\\
 &-2 I[1,2,1,2,0]+I[2,0,1,0,0]-4 M^2 I[2,1,1,0,0]+2 I[2,1,1,0,2]-2 I[2,1,1,2,0]\\
 &+4 M^4 I[2,2,1,0,0]-4 M^2 I[2,2,1,0,2]+I[2,2,1,0,4]+4 M^2 I[2,2,1,2,0]\\
 &\left.-2 I[2,2,1,2,2]+I[2,2,1,4,0]\right),\\
 D_4&=D_5=D_6=D_7=2 i M^2 \kappa ^2 \lambda ^2 \left(-I[0,2,1,0,0]-I[1,1,1,0,0]+2 M^2 I[1,2,1,0,0]\right.\\
 &\left.-I[1,2,1,0,2]+I[1,2,1,2,0]\right),\\
 D_8&=8 i M^2 \kappa ^2 \lambda ^2 I[1,1,1,0,0],\\
 D_9&=-\frac{1}{8} i \kappa ^2 \lambda ^2 \left(32 M^6 I[1,3,1,0,0]-58 M^4 I[0,3,1,0,0]-16 M^4 I[1,2,1,0,0]\right.\\
 &+8 M^4 I[1,3,1,0,2]-8 M^4 I[1,3,1,2,0]-29 M^2 I[0,2,1,0,0]-14 M^2 I[1,1,1,0,0]\\
 &-8 M^2 I[1,2,1,0,2]+8 M^2 I[1,2,1,2,0]+63 I[0,1,1,0,0]+9 I[1,1,1,0,2]\\
 &\left.-9 I[1,1,1,2,0]\right),\\
 D_{10}&=-\frac{1}{8} i \kappa ^2 \lambda ^2 \left(36 I[0,1,1,0,0]+6 M^2 I[0,2,1,0,0]-18 M^2 I[1,1,1,0,0]+9 I[1,1,1,0,2]\right.\\
 &-9 I[1,1,1,2,0]-32 M^4 I[1,2,1,0,0]-12 M^2 I[1,2,1,0,2]+12 M^2 I[1,2,1,2,0]\\
 &\left.+16 M^6 I[2,2,1,0,0]+4 M^4 I[2,2,1,0,2]-4 M^4 I[2,2,1,2,0]\right),\\
 D_{11}&=D_{12}=i \kappa ^2 \lambda ^2 \left(12 I[0,1,1,0,0]+4 M^2 I[0,2,1,0,0]+3 I[1,1,1,0,2]-3 I[1,1,1,2,0]\right.\\
 &\left.-8 M^4 I[1,2,1,0,0]-2 M^2 I[1,2,1,0,2]+2 M^2 I[1,2,1,2,0]\right),\\
 D_{13}&=-4 i \kappa ^2 \lambda ^2 \left(2 I[0,1,1,0,0]+2 M^2 I[1,1,1,0,0]+I[1,1,1,0,2]-I[1,1,1,2,0]\right),\\
 D_{14}&=-2 i \kappa ^2 \lambda ^2 \left(2 I[0,1,1,0,0]-2 M^2 I[1,1,1,0,0]+I[1,1,1,0,2]-I[1,1,1,2,0]\right),\\
 D_{15}&=\frac{1}{2} i \kappa ^2 \lambda ^2 \left(4 M^4 I[1,2,1,0,0]-5 M^2 I[0,2,1,0,0]-4 M^2 I[1,1,1,0,0]-4 M^2 I[1,2,1,0,2]\right.\\
 &+4 M^2 I[1,2,1,2,0]+2 I[0,1,1,0,0]+I[0,2,1,0,2]-I[0,2,1,2,0]+I[1,0,1,0,0]\\
 &\left.+2 I[1,1,1,0,2]-2 I[1,1,1,2,0]+I[1,2,1,0,4]-2 I[1,2,1,2,2]+I[1,2,1,4,0]\right),
\end{align*}
the total being
\begin{align}
\nonumber -i\Sigma_{\phi}[p=0]=&\frac{1}{8} i \kappa ^2 \lambda ^2\Big(M^4 (18 I[0,3,1,0,0]+16 I[1,2,1,0,0]-40 I[1,3,1,0,2]\\
\nonumber &+40 I[1,3,1,2,0]-16 I[2,1,1,0,0]-20 I[2,2,1,0,2]+20 I[2,2,1,2,0])\\
 \nonumber&+M^2 (23 I[0,2,1,0,0]+8 I[0,3,1,0,2]-8 I[0,3,1,2,0]\\
 \nonumber&-68 I[1,2,1,0,2]+68 I[1,2,1,2,0]+8 I[1,3,1,0,4]-16 I[1,3,1,2,2]\\
\label{eq:total_integrals}&+8 I[1,3,1,4,0]+4 I[2,0,1,0,0]+8 I[2,1,1,0,2]-8 I[2,1,1,2,0]\\
 \nonumber&+4 I[2,2,1,0,4]-8 I[2,2,1,2,2]+4 I[2,2,1,4,0])+5 I[0,1,1,0,0]\\
 \nonumber&+4 I[0,2,1,0,2]-4 I[0,2,1,2,0]+4 I[1,0,1,0,0]-10 I[1,1,1,0,2]\\
 \nonumber &+10 I[1,1,1,2,0]+4 I[1,2,1,0,4]-8 I[1,2,1,2,2]+4 I[1,2,1,4,0]\Big)
 .
\end{align}
Before substituting the results of the integration in dimensional regularization, it is worth to dwell upon the the degree of divergence of the contributing integrals. The presence of Lifshitz propagators modifies the usual power counting, and if the integrals $I[\a,\b,\g,\r,\s]$ were to be computed with a cutoff regularization, the leading dependence on the cutoff $\LUV$ would be
\begin{align}
\label{eq:dod}
 I[\alpha,\beta,\gamma,\rho,\sigma]\sim \Lambda^{2+2\frac{\s}{3}}\LUV^{6-2\a-2\b-2\g+\r+\frac{\s}{3}}.
\end{align}
From this one can conclude that the divergences in $-i\Sigma_{\phi}[p=0]$ are at worst quadratic, which is an improvement with respect to the quartic divergences that ordinary scalar gravity would give rise to. Still, the dreaded quadratic divergences do not cancel and  sneak back into the theory because of the hard SUSY breaking entailed by the scalar graviton interactions. However, as follows from eq.~\eqref{eq:dod} and the $\kappa^2$ dependence in eq.~\eqref{eq:total_integrals}, these divergences come with factors of $(\frac{\Lambda}{M_P})^2\Lambda^{2\frac{\s}{3}}$, so that they are strongly suppressed for $\Lambda\ll M_P$. Expression (\ref{eq:dod}) is deduced for $z=3$, 
while higher $z$ lead to  a higher power of $\Lambda$.

A similar suppression holds for the results in the limit of small $\Lambda$ in dimensional regularization, including the finite parts. Using the analytic formulae in \S~\ref{app:integrals}, we obtain the following expression valid for $x=0$ in the limit of small $\Lambda$:
\begin{align*}
 -i\Sigma_{\phi}[p=0]&\sim\frac{ i \kappa ^2 \lambda ^2 \Lambda ^2 M^2}{256\pi^4}\Big\{\frac{18 }{ \epsilon ^2}-\frac{12}{\epsilon} \left(-3+3 \gamma+\text{Log}\left[\frac{M^4 \Lambda ^2}{8 \pi ^3 \mu ^6}\right]\right)\\
 &+ 80-\pi ^2+72 \text{Log}[2 \pi ]+36 \left(\gamma^2+\text{Log}[2]^2-2 \gamma (1+\text{Log}[2 \pi ])\right.\\
 &\left.+\text{Log}[\pi ] \text{Log}[4 \pi ]\right)+16 \text{Log}\left[\frac{M^2 \Lambda }{\mu ^3}\right] \left(-3+3 \gamma+\text{Log}\left[\frac{M^2 \Lambda }{8 \pi ^3 \mu ^3}\right]\right)+O(\Lambda^\frac{10}{3}).
\end{align*}
Despite the fact that some diagrams are IR divergent in the limit $x\rightarrow0$ (those involving the integral $I[1,1,1,0,0]$, see appendix \ref{app:integrals}), the final result is IR safe as expected for physical observables.


\section{Conclusions\label{sec:conclusions}}

The main conclusion of our paper is that the combination of a supersymmetric 
matter sector and Ho\v rava-Lifshitz gravity gives rise to interesting models, 
in which both Lorentz violation and SUSY breaking have a common origin and are controlled by a single 
dimensionless ratio, 
\be
\frac{\Delta c}{c} \propto \frac{\LHL^2}{M_P^2};~~~ m^2_{\rm ~SUSY~breaking} \propto \LUV^2 \times  \frac{\LHL^2}{M_P^2}.
\ee
The consideration of a very large ultraviolet scale, $\LUV \sim M_P$, and the requirement of a natural resolution to the 
gauge hierarchy problem then imply
\be
\LHL \sim {\rm weak ~scale}.
\ee
Supersymmetry in the matter sector also serves as a good argument for explaining why 
LV {\em without} the involvement of gravity is pushed to irrelevant operators. 

To demonstrate our main point we took the simplest supersymmetric Wess-Zumino model with two chiral superfields, 
heavy and light, and coupled it to scalar gravity with Lifshitz scaling. A direct calculation  in dimensional regularization of  two-loop quantum corrections to the light scalar's mass due to the gravitational interactions and the  heavy fields
revealed  the universality of the $\LHL^2/M_P^2$ suppression. While the UV sensitivity of light masses 
remains, it is rendered harmless by the wide separation between the HL and Planck scales. 

Our approach puts  the gravitational force in a completely separate category
from the rest of the interactions: it is not supersymmetric, violates Lorentz symmetry maximally, and acquires a Lifshitz scaling at relatively low energies ({\em e.g.} weak scale). If out of this one may eventually build a reliable theory of quantum gravity, it is a 
relatively modest theoretical price to pay.

\section*{Acknowledgements}
Research at the Perimeter Institute is supported in part by the Government of Canada through Industry Canada, and by the Province of Ontario through the Ministry of Research and Information (MRI). CT acknowledges support from the Spanish Government through grant FPA2011-24568.
MP would also like to acknowledge prior collaborative work and discussions with Yanwen Shang, as well as useful exchange of ideas with the participants of the Kavli IPMU focus week on Gravity and Lorentz violations, Tokyo, Japan, Feb 2013.

\newpage
\appendix
\section{Evaluation of 2 loop integrals\label{app:integrals}}

All relevant two-loop diagrams can be written in terms of a family of 2-loop integrals with two heavy ordinary propagators and one Lifshitz propagator. We consider dimensional regularization in $D=4-2\epsilon$ dimensions. The family of integrals is given by
\begin{align*}
 I[\alpha,\beta,\gamma,\rho,\sigma]=\int \frac{d^Dk d^Dl}{(2\pi)^{2D}}\frac{(l_0)^\rho (|\vec{l}|)^\sigma}{[k^2-M^2]^\alpha[(k+l)^2-M^2)]^\beta[(l^2-\Lambda^{-4}|\vec{l}|^6-\delta m^2)]^\gamma}.
\end{align*}
The usual power counting is modified in the presence of propagators of Lifshitz type. If one were to define the integrals by means of a cutoff regularization with cutoff $\Lambda_C$, the leading cutoff dependence would be
\begin{align*}
 I[\alpha,\beta,\gamma,\rho,\sigma]\sim \Lambda^{2+2\frac{\s}{3}}\Lambda_C^{6-2\a-2\b-2\g+\r+\frac{\s}{3}}.
\end{align*}
Thus the degree of divergence  is lowered with respect to the one that would be obtained with ordinary propagators.

In the case in which all parameters $\a,\b,\g,\r,\s$ are greater than zero, we can reduce these integrals to a single one-dimensional complex integral applying the techniques in ref.~\cite{Smirnov:2006ry} as follows. First, we apply the identities
\begin{align}\label{eq:feynmanpars}
 \frac{1}{A^\alpha B^\beta}=\frac{\Gamma[\alpha+\beta]}{\Gamma[\alpha]\Gamma[\beta]}\int_0^1 dx \frac{x^{\alpha-1}(1-x)^{\beta-1}}{[xA+(1-x)B]^{\alpha+\beta}},\\
 \nonumber\int \frac{d^Dk}{(2\pi)^D} \frac{1}{[k^2-M^2]^\lambda}=\frac{i(-1)^\lambda}{2^D\pi^{D/2}}\frac{\Gamma\left[\lambda-\frac{D}{2}\right]}{\Gamma[\lambda]}(M^2)^{\frac{D}{2}-\lambda}
\end{align}
to perform the integration in $k$. Applying eq.~\eqref{eq:feynmanpars} again one is left with the following integral:
\begin{align}
 &\nonumber\int_0^1 d\xi\, d\eta\,\int d^Dl\, \frac{i(-1)^{\alpha+\beta}\Gamma[\a+\b+\g-D/2]\xi^{\a-1}(1-\xi)^{\b-1}\eta^{\g-1}
 (1-\eta)^{\a+\b-D/2-1}l_0^{\rho}|\vec{l}|^\sigma}{2^{2D}\pi^{3D/2}\Gamma[\alpha]\Gamma[\b]\Gamma[\g]([\eta-(1-\eta)\xi(1-\xi)][l^2-\tilde\Lambda^{-4}|\vec{l}|^6-\tilde M^2])^{\a+\b+\g-D/2}},\\
 &\nonumber\tilde \Lambda=\Lambda\left(\frac{\eta}{\eta-(1-\eta)\xi(1-\xi)}\right)^{-\frac{1}{4}},\\
 &\tilde M^2=\frac{M^2(\eta x-(1-\eta))}{\eta-(1-\eta)\xi(1-\xi)},\label{eq:int2}
\end{align}
where we defined
\begin{align*}
 x\equiv\frac{\delta m^2}{M^2}.
\end{align*}
In order to perform the $l$ integral, we use the propagator representation
\begin{align}
\label{eq:Schwingerprop}
 \frac{1}{A^c}=\frac{i^c\lambda^{c-1}}{\Gamma[c]}\int_0^\infty d\lambda e^{-i\lambda A}.
\end{align}
The $l^0$ integral can be computed after a proper contour deformation, and is given by
\begin{align*}
 \int_{-\infty}^\infty dl_0 l_0^\rho e^{-i\lambda l_0^2}=\!-\frac{(i\!-\!1)^{\rho+1}}{2^\frac{\rho+1}{2}}\int_{\!-\infty}^\infty dy y^\rho e^{-\lambda y^2}=-\frac{(-1)^{3/4}}{2}(1+(-1)^\rho)
 e^{\frac{3i\pi\rho}{4}}\lambda^{-1/2-\rho/2}\Gamma\left[\frac{1+\rho}{2}\right].
\end{align*}
The integral along $d^3\vec{l}$ can be expressed in terms of hypergeometric functions. We are interested in threshold effects from very heavy fields mediated by the scalar gravity
interactions, which explicitly break supersymmetry and hence violate the nonrenormalization of the superpotential. For this reason we are interested in the limit of $M\gg\Lambda$ 
(very heavy thresholds). In this limit the dominant contribution to the integral over $d^3\vec{l}$ is
\begin{align*}
&\frac{2\pi^{\frac{D-1}{2}}}{\Gamma[\frac{D-1}{2}]}\int_0^\infty d|\vec{l}||\vec{l}|^{D-2+\sigma}e^{i\lambda(|\vec{l}|^2+\tilde\Lambda^{-4}|\vec{l}|^6+\tilde M^2)}=\\
&=-\frac{1}{6} (-1)^{11/12} \lambda^{\frac{1}{6} (-D-\sigma+1)} \Lambda^{\frac{2}{3} (D+\sigma-1)} \Gamma \left[\frac{1}{6} (D+\sigma-1)\right] e^{\frac{1}{12} i \pi  (D+\sigma)+i \lambda M^2}
+O(\Lambda^{\frac{2}{3}+D+\sigma}).
\end{align*}
In this limit one may perform the integral over $\lambda$; also, it is useful to rewrite the part of the denominator in eq.~\eqref{eq:int2} involving $\xi,\eta$ as a product of factors, by using the
Mellin-Barnes identity
\begin{align*}
\frac{1}{(A+B)^c}=\frac{1}{2\pi i\Gamma[c]}\int_{-i\infty}^{i\infty}dz\frac{A^z\Gamma[c+z]\Gamma[-z]}{B^{c+z}}.
\end{align*}
Here the contour is taken between the left and right handed poles of the Gamma functions  --the left poles are those corresponding to the factors $\Gamma[z+\dots]$, and the right poles 
 those of the factors $\Gamma[-z+\dots]$. In this way one gets the following representation of the integral $I[\a,\b,\g,\rho,\sigma]$:
\begin{align*}
 &I[\alpha,\beta,\rho,\sigma]= \int_{-i\infty}^{i\infty}dz\int_0^1 d\xi\, d\eta\,\frac{(-1)^{\g}i^{\r-D}
 (1+(-1)^\rho)
 \Gamma[\frac{1}{6}(D+\s-1)]}{2^{2D}\pi^{D+\frac{1}{2}}6\Gamma[\a]\Gamma[\b]
 \Gamma[\g]\Gamma[\frac{D-1}{2}]}\Gamma[-z] \Gamma\left[\frac{1+\r}{2}+z\right]\times\\
 &\times\Gamma\left[-\frac{1}{3}+\a+
 \b+\g-\frac{2D}{3}-\frac{\rho}{2}-\frac{\sigma}{6}\right]\Lambda^\frac{2(D-1+\s)}{3} M^{\frac{1}{3}(2-6\a-6\b-6\g+4D+3\r+\s)}\times\\
 &\times\frac{\xi^{\a-1}(1-\xi)^{\b-1}\eta^{\g-1+z-\frac{1}{6}(D+\s-1)}(1-\eta)^{\a+\b-\frac{D}{2}-1}(x\eta-(1-\eta))^{\frac{1}{6}(2-6\a-6\b-6\g+4D+3\rho+\s)}}{[-(1-\eta)\xi(1-\xi)]
 ^{\frac{1+\r}{2}+z}}.
\end{align*}
The integral in the parameters $\xi,\eta$ can be expressed in terms of Gamma functions, leaving only a one dimensional contour integral:
\begin{align*}
 &I[\alpha,\beta,\rho,\sigma]= \int_{-i\infty}^{i\infty}dz\,\frac{i^{1-\r}(-1)^{\a+\b+\g}4^{-1-D}(1+(-1)^\rho)
 \Gamma[\frac{1}{6}(D+\s\!-\!1)]}{3\pi^{D+\frac{3}{2}}\Gamma[\a]\Gamma[\b]
 \Gamma[\g]\Gamma[\frac{D-1}{2}]\Gamma[\a+\b-\rho-2z-1]}\Gamma\left[\!-\frac{1}{2}+\a\!-\frac{\rho}{2}\!-z\right]\times\\
 &\times\Gamma\left[-\frac{1}{2}+\b-\frac{\rho}{2}-z\right]\Gamma\left[-\frac{1}{2}+\a+\b-\frac{D}{2}-\frac{\rho}{2}-z\right]\Gamma[-z]\Gamma\left[\frac{1+\r}{2}+z\right] \times\\
 &\times\Gamma\left[\frac{1}{6}-\frac{D}{6}+\g-\frac{\sigma}{6}+z\right]\Lambda^\frac{2(D-1+\s)}{3} M^{\frac{1}{3}(2-6\a-6\b-6\g+4D+3\r+\s)}
 x^{\frac{1}{6}(-1+D-6\g+\sigma-6z)}.
\end{align*}
Again, the contour integral runs between the right and left poles of the Gamma functions. The singularities in $D=4$ come from either z-independent Gamma functions or when the integration contour is pinched between poles that approach
as $D\rightarrow 4$. The latter can be isolated by appropriately deforming the integration contour, so that $I[\a,\b,\g,\rho,\sigma]$ can be expressed as a sum over residues over some of 
the poles pinching the integration contour plus another line integral free of singularities as $D\rightarrow 4$, for which the latter limit may be safely taken. 

As an example, let's evaluate $I[1,1,1,0,0]$. From eq.~\eqref{eq:masterint}, one can see that the left-handed poles in $z$  of the Gamma functions sit at $\{-\frac{1}{2}-k,\,-\frac{1}{2}-\frac{\epsilon}{3}-k\},\,k=0,1,2,\dots$, while the right handed poles are
at  $\{\frac{1}{2}+k,\,-\frac{1}{2}+\epsilon+k\},\,k=0,1,2,\dots$. Thus there is a pinch singularity when $D\rightarrow4$ ($\epsilon\rightarrow0$) as the contour is trapped between the poles at $z=-\frac{1}{2}$ and 
$z=-\frac{1}{2}+\epsilon$. The contour can be deformed as the sum of a vertical line to the left of the $z=-\frac{1}{2}-\frac{\epsilon}{3}$ pole plus a sum of residues over
$z=-\frac{1}{2}-\frac{\epsilon}{3}$ and $z=-\frac{1}{2}$. The line integral is finite for $D\rightarrow4$ and can be evaluated closing the contour on the left and summing over 
the residues of the poles inside the contour; the sum
converges very quickly and we choose to approximate it by the first terms corresponding to the poles closer to the origin, which is equivalent to an expansion in $x$; in this case, the first term is already of order $x^2$ and we will neglect it. We are left just with the residues of the poles giving rise to the singularity, that is
\begin{align*}
&I[1,1,1,0,0]= \Lambda^2\left\{\frac{1}{512 \pi ^4 \epsilon ^2}-\frac{1}{1536 \pi ^4 \epsilon }\left(\log \left(\frac{\Lambda ^4 M^8 x^4}{16384 \pi ^6}\right)+6 (\gamma -2)\right)\right.\\
&\!+\!\frac{1}{27648\pi^4}\Big[48 \log (\Lambda ) \left(\log \left(\frac{\Lambda  M^4 x^2}{128 \pi ^3}\right)-6\right)+96 \log (M) \left(\log \left(\frac{x^2}{128 \pi ^3}\right)+3 \gamma -6\right)\\
&+192 \log ^2(M)+144 \gamma  \log (\Lambda  x)+12 \log (x) (\log (x)-4 (3+\log (16)+3 \log (\pi )))-19 \pi ^2\\
&+12 \left(30+9 \log ^2(\pi )+6 (6\right.\\
&\left.+\log (128)) \log (\pi )+\log (2) (60+37 \log (2))\right)+36 \gamma  (3 \gamma -2 (6+\log (128)+3 \log (\pi )))\Big]
\Big\},\end{align*}
where $\gamma$ is the Euler constant. In a similar way, results for all the integrals needed for the computation follow. Where appropriate, we have computed line integrals which are finite at $D=4$ by closing contours and summing over residues.
These sums are dominated by the residues of the poles closest to the origin, and we give the corresponding analytic expressions --in fact, the residues of these poles 
involve powers of $x=\delta m^2/M^2$ that increase with the distance to the origin, so that the formulae below correspond to the lowest terms in an $x$ expansion. Thus, in the previous formula and the ones that follow, equalities are understood up to higher orders in $\Lambda$ and $x$.
\begin{align*}
&I[1,1,1,2,0]= \Lambda^2 M^2\Big\{\frac{x+12}{1024 \pi ^4 \epsilon ^2}+\frac{1}{1536 \pi ^4 \epsilon}\Big[\!-24 \log \left(\Lambda  M^2\right)-2 x \log \left(\Lambda  M^2 x\right)-3 \gamma  (x+12)\\
&+x (5+\log (128)+3 \log (\pi ))+60+36 \log (2 \pi )\Big]+\frac{1}{55296 \pi ^4}\Big[576 (3 \gamma -5) \log \left(\Lambda  M^2\right)\end{align*}\begin{align*}
&+12 \left(-2 x \log (x) \left(-4 \log \left(\Lambda  M^2\right)+7+\log (256)\right)+4 \log \left(\Lambda  M^2\right) \left((x+12) \log \left(\Lambda  M^2\right)\right.\right.
\\
&\left.\left.-x (5\!+\!\log (128))\!-\!12 \log \left(8 \pi ^3\right)\right)+12 x (\gamma \!-\!\log (\pi )) \log \left(\Lambda  M^2 x\right)+x \log ^2(x)\right)+108 \gamma ^2 (x+12)\\
&-\pi ^2 (19 x+36)+12 \left(x \left(32+9 \log ^2(\pi )+6 (5+\log (128)) \log (\pi )+\log (2) (58+37 \log (2))\right)\right.\\
&\left.\!+\!12 \left(38\!+\!9 \log ^2(2)\!+\!30 \log (2 \pi )+3 \log (\pi ) \log \left(64 \pi ^3\right)\right)\right)-72 \gamma  (x (5+\log (128)+3 \log (\pi ))+60\\
&+36 \log (2 \pi ))\Big]\Big\},\\
&I[1,1,1,0,2]= (x M^2\Lambda^{10})^{1/3}\Big\{\frac{\Gamma \left(-\frac{1}{3}\right) \Gamma \left(\frac{5}{6}\right)}{384 \pi ^{9/2} \epsilon }
+\frac{1}{1152 \pi ^{9/2}}\Big[\Gamma \left(-\frac{1}{3}\right) \Gamma \left(\frac{5}{6}\right) \left(\log \left(\frac{256 \pi ^6}{\Lambda ^4 M^8 x}\right)\right.\\
&\left.-\frac{\pi }{\sqrt{3}}-6 \gamma +9\right)\Big]-\frac{(x-84) \Gamma \left(-\frac{7}{6}\right) \Gamma \left(-\frac{1}{3}\right)}{13824 \left(\sqrt[3]{2} \sqrt{3} \pi ^{7/2} \sqrt[3]{x}\right)}
-\frac{x \Gamma \left(-\frac{4}{3}\right) \Gamma \left(\frac{5}{6}\right)}{4608 \pi ^{9/2}}\Big\},\\
&I[1,2,1,4,0]=(\Lambda M)^2\Big\{\frac{3 (x+4)}{1024 \pi ^4 \epsilon ^2}-\frac{1}{1024 \pi ^4 \epsilon }(x+4) \left(\log \left(\frac{\Lambda ^4 M^8}{64 \pi ^6}\right)+6 \gamma -8\right)\\
&+\frac{1}{6144 \pi ^4}\Big[\!-\!256 \log (\Lambda )+4 \left(4 (x+4) \left(\log \left(\frac{\Lambda ^4}{64 \pi ^6}\right)+6 \gamma -8\right) \log (M)+16 (x+4) \log ^2(M)\right.\\
&\left.+\log \left(\frac{\pi ^3}{\Lambda ^2}\right) (-2 (x+4) \log (\Lambda )-6 \gamma  (x+4)+x (8+\log (64)+3 \log (\pi ))+12 \log (4 \pi ))\right),\\
&+36 \gamma ^2 (x+4)-\pi ^2 (x+4)-24 \gamma  (x+4) (4+\log (8))+12 \log (2) (8 x+(x+4) \log (8))\\
&+32 (3 x+2 (7+\log (64)+6 \log (\pi )))\Big]\},\\
&I[1,2,1,0,4]=\frac{\Lambda^\frac{14}{3}}{M^\frac{2}{3}}\Big\{\frac{(x-60) \Gamma \left(-\frac{2}{3}\right)}{13122 \pi ^2 \Gamma \left(\frac{5}{3}\right) \Gamma \left(\frac{11}{3}\right)}
+\frac{x^{2/3} (x+20) \Gamma \left(-\frac{5}{3}\right) \Gamma \left(\frac{7}{6}\right)}{9216 \pi ^{9/2}}\Big\},\\
&I[1,2,1,2,2]=\Lambda^\frac{10}{3}M^\frac{2}{3}\frac{(x-12) \Gamma \left(-\frac{1}{3}\right)}{3888 \pi ^2 \Gamma \left(\frac{1}{3}\right) \Gamma \left(\frac{7}{3}\right)},\\
&I[1,3,1,2,0]= \frac{\Lambda^2}{M^2}\Big\{-\frac{1}{256 \pi ^4 \epsilon }+\frac{\log \left(\frac{\Lambda ^4 M^8}{64 \pi ^6}\right)+6 \gamma -5}{768 \pi ^4}+\frac{x (-6 \log (x)+1+6 \log (4))}{27648 \pi ^4}\Big\},\\
&I[1,3,1,0,2]=\left(\frac{\Lambda }{M}\right)^{10/3}\Big\{-\frac{\sqrt[3]{x} (9 x+80) \Gamma \left(-\frac{4}{3}\right) \Gamma \left(\frac{5}{6}\right)}{138240 \pi ^{9/2}}+\frac{175 (2 x+39) \Gamma \left(\frac{5}{3}\right)}{118098 \pi ^2 \Gamma \left(\frac{10}{3}\right) \Gamma \left(\frac{16}{3}\right)}\Big\},\\
&I[2,2,1,2,0]= \frac{\Lambda^2}{M^2}\Big\{-\frac{x-6 x \log (2)+3 x \log (x)+36}{13824 \pi ^4}\Big\},\\
&I[2,2,1,0,2]=\left(\frac{\Lambda}{M}\right)^\frac{10}{3}\Big\{-\frac{\sqrt[3]{x} (3 x+20) \Gamma \left(-\frac{4}{3}\right) \Gamma \left(\frac{5}{6}\right)}{34560 \pi ^{9/2}}
+\frac{35 (25 x+312) \Gamma \left(\frac{2}{3}\right)}{354294 \pi ^2 \Gamma \left(\frac{10}{3}\right) \Gamma \left(\frac{16}{3}\right)}\Big\},\\
&I[1,2,1,2,0]= \Lambda^2\Big\{\frac{3}{512 \pi ^4 \epsilon ^2}+\frac{\log \left(\frac{64 \pi ^6}{\Lambda ^4 M^8}\right)-6 \gamma +6}{512 \pi ^4 \epsilon }\\
&+\frac{1}{3072 \pi^4}\Big[16 \log \left(\Lambda  M^2\right) \left(\log \left(\Lambda  M^2\right)+3 \gamma -3-3 \log (2 \pi )\right)-\pi ^2+36 \gamma  (\gamma -2 (1+\log (2\pi)))\end{align*}\begin{align*}
&+36 (2+\log (2 \pi ) (2+\log (2 \pi )))\Big]+\frac{x \left(\log \left(\frac{x}{4}\right)-1\right)}{1536 \pi ^4}\Big\},\\
&I[1,2,1,0,2]= \left(
\frac{\Lambda^{10}}{M^4}\right)^\frac{1}{3}\Big\{-\frac{10 (x+42) \Gamma \left(-\frac{4}{3}\right)}{59049 \pi ^2 \Gamma \left(\frac{7}{3}\right) \Gamma \left(\frac{13}{3}\right)}+\frac{\sqrt[3]{x} (x+16) \Gamma \left(-\frac{4}{3}\right) \Gamma \left(\frac{5}{6}\right)}{9216 \pi ^{9/2}}\Big\},\\
&I[1,3,1,4,0]=\Lambda^2\Big\{-\frac{1}{512 \pi ^4 \epsilon }+\frac{\log \left(\frac{\Lambda ^4 M^8}{64 \pi ^6}\right)+6 \gamma -6}{1536 \pi ^4}+\frac{x \log \left(\frac{4}{x}\right)}{1536 \pi ^4}\Big\},\\
&I[1,3,1,0,4]=\frac{\Lambda^\frac{14}{3}}{M^\frac{8}{3}}\Big\{\frac{5 (7 x-132) \Gamma \left(\frac{4}{3}\right)}{39366 \pi ^2 \Gamma \left(\frac{8}{3}\right) \Gamma \left(\frac{14}{3}\right)}
 -\frac{x^{2/3} (9 x+100) \Gamma \left(-\frac{5}{3}\right) \Gamma \left(\frac{7}{6}\right)}{138240 \pi ^{9/2}}\Big\},\\
 &I[1,3,1,2,2]=\frac{\Lambda^\frac{10}{3}}{M^\frac{4}{3}}\Big\{\frac{(5 x+84) \Gamma \left(-\frac{4}{3}\right) \Gamma \left(\frac{5}{3}\right)}{11664 \sqrt{3} \pi ^3 \Gamma \left(\frac{10}{3}\right)}
 -\frac{x^{4/3} \Gamma \left(-\frac{4}{3}\right) \Gamma \left(\frac{5}{6}\right)}{4608 \pi ^{9/2}}\Big\},\\
 &I[2,2,1,4,0]=\Lambda^2\Big\{\frac{3}{256 \pi ^4 \epsilon ^2}+\frac{\log \left(\frac{64 \pi ^6}{\Lambda ^4 M^8}\right)-6 \gamma +5}{256 \pi ^4 \epsilon }
 +\frac{1}{1536 \pi ^4}\Big[40 \log \left(\frac{4}{\Lambda }\right)\\
 &+16 \log (\Lambda ) (\log (\Lambda )+3 \gamma -3 \log (2 \pi ))+16 (4 \log (\Lambda )+6 \gamma -5-6 \log (2 \pi )) \log (M)\\
 &+64 \log ^2(M)-\pi ^2+4 (15+\log (2) (\log (512)-5)+3 \log (\pi ) (5+\log (64)+3 \log (\pi )))\\
 &+12 \gamma  (3 \gamma -5-6 \log (2 \pi ))\Big]-\frac{x}{768 \pi ^4}\Big\}, \\
 &I[2,2,1,0,4]=\frac{\Lambda^\frac{14}{3}}{M^\frac{8}{3}}\Big\{\frac{10 (33-4 x) \Gamma \left(-\frac{2}{3}\right)}{177147 \pi ^2 \Gamma \left(\frac{8}{3}\right) \Gamma \left(\frac{14}{3}\right)}
 -\frac{x^{2/3} (3 x+25) \Gamma \left(-\frac{5}{3}\right) \Gamma \left(\frac{7}{6}\right)}{34560 \pi ^{9/2}}\Big\},\\
 &I[2,2,1,2,2]=\frac{\Lambda^\frac{10}{3}}{M^\frac{4}{3}}\Big\{\frac{(x-21) \Gamma \left(\frac{11}{6}\right)}{2430 \sqrt[3]{2} \pi ^{5/2} \Gamma \left(\frac{10}{3}\right)}
 -\frac{x^{4/3} \Gamma \left(-\frac{4}{3}\right) \Gamma \left(\frac{5}{6}\right)}{4608 \pi ^{9/2}}\Big\}.
\end{align*}
Except for $I[1,1,1,0,0]$, the previous integrals have a well defined limit when $x\rightarrow0$.  In the case $\alpha=0$ (or equivalently $\beta=0$), with the rest of parameters staying positive, the two-loop integral factorizes in one-loop integrals, and the $l$ integral can be performed 
using eq.~\eqref{eq:Schwingerprop} as before. Doing a series expansion in $\Lambda$ and keeping the lowest order, one gets
\begin{align*}
 I[0,\b,\g,\r,\s]=&I[\b,0,\g,\r,\s]=\frac{i^{2\g+\r}(1+(-1)^\rho)}{3\cdot2^{2+D}\pi^{D/2}\Gamma[\b]\Gamma[\g]}\Gamma\Big[\b\!-\!\frac{D}{2}\Big]\Gamma\Big[\frac{1+\rho}{2}\Big]\Gamma\Big[\frac{1}{6}(\!-\!1+D+\s)\Big]\times\\
 &\times\Gamma\Big[\frac{1}{6}(-2-D+6\g-3\r-\s)\Big]\Lambda^{\frac{2}{3}(-1+D+\s)}M^{\frac{1}{3}(D-2\b)}\delta m^{\frac{1}{3}(2+D-6\g+3\r+\s)}.
\end{align*}
In the case $x=0$, the $l$ integral is of the tadpole type and it vanishes in dimensional regularization, so that
\begin{align*}
  I[0,\b,\g,\r,\s]|_{x=0}=0.
\end{align*}

\newpage
\bibliographystyle{h-physrev}
\bibliography{bib-WZ_Horava}

\end{document}